# Thomas Hinsley Astbury: from an English market town schoolroom to the internal constitution of the stars

**Jeremy Shears**

## Abstract

T.H. Astbury (1858-1922) was for many years the much-respected headmaster of a boys' junior school in the English market town of Wallingford. By night he was a dedicated amateur astronomer who enjoyed observing meteors, variable stars and many other objects. He began to search few new variable stars, his first discovery being the bright Cepheid variable, RT Aurigae. This, along with his discovery of 4 other variable stars, brought him to attention of some of the most famous professional astronomers of the age, including Herbert Hall Turner, Frank Dyson and Arthur Eddington.

## Introduction

One afternoon in November 2011, whilst researching another topic in the Royal Astronomical Society library, I was perusing the May 1923 edition of the BAA *Journal* when I came across the following notice: (1)

"*PROPOSED MEMORIAL TO THE LATE T. H. ASTBURY*

The late Mr. T. H. Astbury was a hard-working schoolmaster, who, while paying full attention to his ordinary duties, and acting as organist on Sundays, devoted time and skill to the discovery of variable stars. He was formerly an active member of our Variable Star Section, and with only a binocular and a three-inch telescope has several such discoveries to his credit, among them one of considerable importance, as the subjoined note by Prof. Eddington makes clear. It has occurred to some of us that it would be in every way appropriate if a small brass could be erected to his memory on the walls of the school where he used to teach, viz., the Council School at Wallingford, 'where he rendered good and faithful service from the time of its opening until his retirement'; these last words are quoted from a letter by the Secretary of the Berkshire Education Committee in signifying his entire approval of the project. The estimate from a local firm suggests that the amount required would be about £10, and it is hoped that the number of those who would sympathise with such a scheme may be sufficient to render it unnecessary for any one of them to give more than this."

The notice was placed by Prof. A.S. Eddington, Prof H.H. Turner, Col. E.E. Markwick and Mr. A.N. Brown. I was surprised by two aspects of the notice. Firstly, that I had never heard of Thomas Hinsley Astbury, in spite of my being an active variable star observer who also has an interest in the history of the BAA Variable Star Section (VSS). The second surprise was that two of the most famous professional astronomers of the day had apparently leant their weight to an appeal to memorialise a schoolteacher and amateur astronomer. Arthur Stanley Eddington (1882-1944) had been Plumian Professor of Astronomy and Experimental Philosophy at Cambridge University since 1913. In 1914 he was elected a Fellow of the Royal Society and won its Royal Medal in 1918 (2). Herbert Hall Turner, FRS (1861-1930) was Savilian Professor of Astronomy at Oxford. Ernest Elliot Markwick (1853-1925) had previously served as Director of the BAA VSS 1899-1909 and as BAA President 1912-1914 (3). A.N. Brown was at the time Secretary of the BAA VSS.

So who was Astbury? And what was his contribution to variable star astronomy that had brought him to the attention of these illustrious scientists? There and then, I resolved to find out more about Astbury, his life and his contribution to variable star observing with the aim of bringing this long forgotten amateur astronomer to the attention of current members of the Association. The present paper is the result of this research.

**The Wallingford schoolmaster**

Thomas Hinsley Astbury (1858-1922; Figure 1) was born at Shifnal in east Shropshire. His father, Joseph Astbury, was an Iron Works Foreman (4), this region being one of the centres of England’s burgeoning iron industry. His mother, Hannah Hinsley, from whom Thomas obtained his middle name, was a dealer in drapery. Joseph and Hannah were married in Hannah’s home town of Madeley, Shropshire (5). The marriage resulted in seven children and the family at various times lived in Shifnal, Wellington (6), also in Shropshire, and Handsworth, West Bromwich, in connexion with Joseph’s employment (7).

Astbury became a schoolteacher by profession (8) and from 1883 taught in the Oxfordshire market town of Wallingford (until boundary changes in 1974, Wallingford was located in the county of Berkshire). Wallingford is situated about 20 km south-east of Oxford and a similar distance north-west of Reading. In 1886 he married Emily Wallis Naish in Wallingford (9). They had 2 daughters (10) and lived at Croft Villas, Wallingford (11). For the majority of his teaching career he taught at Kinecroft School, a primary school in Wallingford (Figure 2) catering for both girls and boys and where he became headmaster. A photograph of Astbury with some of the other schoolteachers is shown in Figure 3.

However, in the first decade of the 1900s it was recognised that the school was outgrowing its premises and plans were soon drawn up for a new school for Junior boys. Thus the Wallingford Council Boys School (12) (Figures 4 and 5) was opened on 4 April 1910, with Astbury as Headmaster (13). The school was built by local company Messrs Boshers, Sons & Co (14) for the sum of £2256 15s 3d and comprised a hall surrounded by 6 classrooms. As with many new buildings, it was not without its teething problems, which Astbury had to deal with, noting (15) for example that “the flushing apparatus at the boys’ offices still does not act and the place is getting very offensive again”. There was also a damp wall and several draughty window frames, not to mention one or two leaks. A further building was added in 1912 which comprised a Cooking and Manual Centre. This was used by the boys for woodwork classes for half the week (Figure 6) and by girls from another school in the town for cookery lessons during the other half.

Astbury remained justifiably proud of the new Wallingford Council School for Boys that he had established and an inspection in March 1911 was complementary about the quality of the education, with some improvements suggested:

"The boys are in excellent discipline, a pleasant tone prevails, the work is carried on with great care, and there are many meritorious points in the school...As a rule the fundamental subjects are effectively taught, but in the reading lessons greater stress should be laid on subject matter and the writing of the highest class should be better".

The Punishment Book for 1910 makes an interesting read. Misdemeanours recorded included: "laughing in class" (1 stripe of the ruler or cane), "writing filthy words" (2 stripes), "letting down bicycle tyres" (2 stripes), "throwing milk on a boy's book" (3 stripes) and "running home from school on Wednesday afternoon and Thursday morning and rudeness and insubordination" (4 stripes).

A little over 4 years after it opened, the First World War started and classes became increasingly disrupted by the absence of teachers on active service (15). Desperate times called for desperate measures, thus by 1917, with several teachers away, Astbury found himself teaching classes of over 100 boys. The boys also played their part in the war effort. In September 1917 when they returned from 5 weeks of summer holidays, they started the collection of horse chestnuts for the Ministry of Munitions of War. The conkers were used to produce acetone which was used in the manufacture of ordnance (16).

Astbury was much admired and respected as headmaster of the school he had created. He eventually retired as headmaster in 1920 due to ill health.

**Astbury and the BAA**

Quite when Astbury's interest in astronomy began is not recorded, but by 1898 he was evidently sufficiently interested in the subject to be elected as a member of the BAA on 23 February (17), having been proposed by Miss Elizabeth Brown (1830-1899) and seconded by E. Bruce Mackay (18). Brown was a well-known figure during the first decade of the fledgling BAA, having been involved in establishing the Association in 1890 (19). She was an active solar observer and was immediately appointed as Director of the Solar Section, having held a similar position in the Liverpool Astronomical Society. Apart from solar work, she was also active in the Variable Star, Lunar and Star Colour sections of the Association. Shortly after Brown's death on 5 March 1899, Astbury wrote a letter (20) to the *English Mechanic* putting on record that "he wished to express his sincere regret for the loss sustained by her death. He would always consider it one of his greatest privileges to have been associated in some degree with her work". Drawings by Astbury of the Great Sunspot Group of September 1898 which he submitted to the BAA Solar Section are shown in Figure 7.

Astbury's first contribution to the BAA *Journal* was a letter (21) concerning an unusual observation he made on the evening of 27 March 1899: "At 8.10…I noticed, some distance to the left of the moon (full moon, 28d 18h 19m) and at an equal elevation, what appeared be another moon shining through a thick haze. Immediately afterwards another, precisely similar, was seen to the right of the moon". He described these as "mock-moons", by analogy to "mock-suns" which are sometimes observed. Later on in the evening he observed a lunar halo. A mock-moon, sometimes called a "moon dog" or *paraselene*, is a bright circular spot on a lunar halo and is an atmospheric phenomenon caused by the refraction of moonlight by hexagonal ice crystals in cirrus or cirrostratus clouds. They appear to the left and right of the moon 22° or more distant. By contrast to mock-suns, or sun dogs, they are infrequently seen because to be produced the moon must be bright and therefore full or nearly full.

Throughout his astronomical career, Astbury used nothing larger than a 3¼ inch (8.3 cm) Wray refractor (22), although many of his observations were made with the naked eye or binoculars (23). He was also an early experimenter in the use of photography to record meteors, but was thwarted by the relatively slow emulsions of the age (24). In 1904 he wrote a letter (25) to the *English Mechanic* confessing that in spite of many attempts during the main meteor showers, he had never succeeded in recording any, although there were plenty of star trails on his plates. In spite of this, he provided advice for readers of the same publication (26) in advance of the 1904 Perseids, including the suggested length of exposure, the type of photographic plates to use and the method of developing them. He was more successful with visual meteor observation and his results are included in the BAA Meteor Section reports from 1898 to 1904 (27). He was fortunate enough in August 1899 to see a fine Perseid equal to the brightness of Sirius which left a train that remained visible for 12 seconds (28). He made this observation from Shifnal, Shropshire. According to W.F. Denning (1848-1931), then Director of the BAA Meteor Section, the same meteor was observed by Rev. S.J. Johnson at Bridport, Dorset (29), and Rev. T.E.R. Phillips (1868-1942) at Yeovil, Somerset, who said it was equal to Venus in brightness. Astbury was active in the Meteor Section until around 1905, when the Section essentially collapsed (30), but even after that he continued to send observations to Denning who particularly valued their accuracy. In some cases, Denning was able to combine Astbury's data on a particular meteor with those of other observers and thus calculate via triangulation its trajectory through the atmosphere (31).

Other astronomical objects and phenomena that Astbury observed included the Sun, as mentioned previously, the zodiacal light (32), aurorae (33) and comets, including Halley's Comet in 1910 (34). He also observed the occultation of Saturn by the Moon on 3 September 1900 using his Wray refractor x 80. Just before the occultation he observed an unusual "prominence" or protuberance near the rings, the origin of which he could not explain (35). Four years later, and still perplexed by his observation, he wrote a further letter to the *Journal* (36), wondering whether a thin lunar atmosphere might have caused differential refraction as Saturn approach the limb. Of course we now know that the moon's atmosphere is so tenuous that it could not cause such effects and what he saw. What he saw, whether real or illusory, remains a mystery (37).

Apart from his meteor and solar work, other observations were generally more of a casual nature, rather than being part of a sustained programme of study. However, it was of course the variable stars that eventually captured his imagination and after 1905 this branch of astronomy essentially dominated his observational work, resulting in his projection to fame as a discoverer of several new variables. Thus it is to his variable star observations that we shall now turn our attention.

## Variable star observing and the discovery of RT Aurigae

Astbury might have begun making variable star observations in earnest with the appearance of Nova Persei in February 1901, as his observations of this object are the earliest that can be traced in the published record (38). He continued to report estimates of the nova's brightness to the BAA VSS through to at least March 1904, by which time the star had faded to about magnitude 10 (Figure 8). After the initial excitement and the associated deluge of observations following the appearance of the nova, only 3 observers continued to report estimates to VSS Director Col. E.E. Markwick during 1903 and 1904 (39): Astbury, Markwick and Charles Lewis Brook (40) (1855 – 1939; Figure 9). Sadly, it is still the case nowadays that most people tend to lose interest and stop observing novae after the initial fade.

According to Markwick (41), Astbury joined the VSS in about 1903, although he had clearly been submitting observations of Nova Persei to him before that. He made observations of a range of variables and some are included in the VSS Memoirs covering the intervals 1900-1904 and 1905-1909 (42). By contrast the subsequent Memoir covering the years 1910 to 1914 contained no observations by him (43). Rather than making routine estimates of known variables, as we shall soon see, Astbury turned to seeking out new variables.

The VSS was formed in 1890 and is today the longest established organisation for the observation of variable stars in the world. The first Director, Markwick's predecessor, was John Ellard Gore FRAS, MRIA (1845-1910), an Irish amateur astronomer and prolific author of popular astronomy books. In addition to collecting observations of known variable stars, Gore had increasingly focussed the work of the Section on a nova search programme and observations of stars which were suspected of being variable. By their nature such programmes generally yield negative results (and a few false alarms in the case of the nova programme). As Toone points out (44), this was probably a strategic mistake as many variable star observers, especially new ones, like actually to see the variable star and detect obvious variations to maintain their interest. By the time Markwick became Director, membership was flagging and the Section was in the doldrums. Markwick brought his organisational and motivational capabilities that he had honed during years of military service to bear on the problem. Thus he set about reinvigorating the section and encouraged cooperation amongst observers on a limited list of variable stars. Through his regular Section circulars and personal letters providing feedback to observers, momentum was regained and membership of the Section began to grow.

Some 3 years after the appearance of Nova Persei, Markwick announced his own "*Plan for watching the Region of the Milky Way for Novae*" under the auspices of the VSS (45). The Milky Way was divided into 6 sections and observers were assigned to these areas with the aim of examining them regularly for novae. Charts were supplied to interested members, but it appears that this project, in contrast to Markwick's other projects, never really took off. Nevertheless, one of the observers who did sign up was Astbury. He was allocated the constellations of Orion, Monoceros, Canis Minor, with part of Taurus and Aurigae (41) and regularly monitored the region with the naked eye, sometimes supplemented with binoculars.

It was whilst engaged in the nova search programme that Astbury discovered his first variable star. In March 1905 he suspected the fifth magnitude star 48 Aurigae of variability. In his own words, from his BAA *Journal* paper "A New Naked-eye Variable": (46)

"As a member of the Variable Star Section I have been asked by Col. Markwick to watch a region of the Milky Way with the idea of detecting the possible appearance of a 'nova'. During these watches my interest was aroused by the star 48 Aurigae….., but until March 7$^{th}$ I was undecided whether the variation suspected was the result of a real change of brightness, or was due to imperfection of vision, or of the observing conditions.

Hitherto observations had been made entirely with the naked eye, but the use of an opera glass on the succeeding evenings confirmed my impression so strongly that I felt quite assured as to the reality of the light changes".

Astbury also "photographed the region, allowing the stars to trail, and the results, although very amateurish, are quite sufficient of themselves to prove variability of the star" (46).

Then, as now, variable star estimates require good comparison star sequences with which to compare the star's brightness. At the time, the most reliable sequences were those from the Harvard College Observatory prepared under the direction of Prof. E.C. Pickering (1846-1919). Astbury wrote to Markwick on 11 March, knowing that he had access to Harvard photometry, and received from him an appropriately annotated chart of the region by return.

In spite of being assured of the star's variability, and having initially estimated that the period was a little more than 7 days, he did not want to make an announcement quite yet, confiding to Markwick on 23 March: (41)

"I do not intend to make any announcement at *present*. I do not feel any doubt of the general accuracy of my estimates, but it is very easy to go wrong in such matters when the light changes are confined within narrow limits".

We now know that those "narrow limits" are actually not much more than half a magnitude, which stands testimony to Astbury's abilities as a variable star observer. He had hoped that further observations might be obtained by other VSS members, but Markwick himself was away on Army business at the. Fortunately he had also notified two other observers, J.E. Gore and Arthur Stanley Williams (1861-1938) (47), both of whom sent him their observations. He also informed Pickering at Harvard and Prof. H. H. Turner at Oxford University of the potential discovery – Turner was a great friend and supporter of the VSS over many years. Turner reviewed Astbury's and Williams' initial observations (48) and satisfied himself that the star was indeed variable (49). On 4 April 1905 Turner penned a letter announcing the discovery on Astbury's behalf to the premier journal for such announcements, *Astronomische Nachrichten* (50). It was given the temporary name 47.1905 Aurigae (51).

Subsequently, Arthur Stanley Williams also analysed his and Astbury's estimates of the star between 18 March and 12 April 1905, finding a period 3.8 days with an amplitude of 0.55 magnitudes (52). From the shape of the light curve (Figure 10) he concluded that it was a Cepheid variable (53). He was clearly impressed by Astbury's visual discovery and in a letter to Astbury on 20 April he wrote: (52)

"It is no mean feat to discover the variability of a short-period star like this, in which the whole range of variation barely exceeds half a magnitude! There is very little merit attaching to the discovery of a variable by photography, but I always feel great admiration for a visual discovery of the kind".

Later, 48 Aurigae was given the official variable star name of RT Aurigae. Astbury received praise for his remarkable discovery and was lauded at the BAA meeting of 26 April 1905 where his discovery paper was read. The President, A.C.D. Crommelin (1865-1939) summed it up by stating that: (54)

"[he] thought it was a matter of which the Association might be proud that one of its members had made this discovery. It seemed to be the brightest variable star that had been discovered for something like 36 years, not including novae but only regular variables. There were very few observers that could say that any of the ordinary naked-eye stars were varying to the extent of half a magnitude. It was a matter that warranted very precise observations, and very careful noting and collating of one's observations….he would ask those present to return their thanks to Mr. Astbury for his paper, and to congratulate him on his brilliant discovery."

Over the next 2 years, Astbury collected further observations of his new Cepheid and prepared a revised light curve, based on a further improvement of the comparison star sequence (55) and his result is shown in Figure 11.

**The search for new variable stars continues**

Spurred on by his discovery of RT Aur, Astbury continued his search for new variables by monitoring 30 stars which he had suspected of variability from his earlier observations (56), but he was not successful in confirming any to be variable. Thus after two years, in 1907, he decided to change his tactics. His new approach was based on a procedure outlined in an 1899 article in the scientific periodical *Knowledge* by J.E. Gore to determine whether any stars in a particular field were variable. Astbury described the method thus: (56)

“Decide upon an easily-recognised group of stars falling within the field of the binocular, and, with the aid of a lantern (preferably with a yellow shade), sketch the positions and magnitudes of the stars as accurately as possible. Now carefully number the stars in order of brightness, beginning with the brightest. In the case of two stars whose difference is considerable drop one or more numbers. For example, 1, 2, 4, 7, 8 = 9 would signify that 2 was perceptibly less than 1, 4 obviously less than 2, 7 much less that 4…….”

Thus, after a time it would become evident if one of the stars had brightened or faded. Astbury commented that "I do not keep any record of these observations until variability is discovered, but a written record would be of the greatest service afterwards when calculating the elements of the variation". He said that it was important to become thoroughly familiar with a particular field and only then should new fields be added: “Experience has proved the possibility of examining a dozen or more groups in the course of half an hour with sufficient care to detect any change”. Almost as if it were an afterthought he commented: (56) “Finally, I would strongly emphasise the necessity for persistent effort”

Fortunately Astbury’s efforts were persistent and his next discovery, an Algol-type eclipsing binary in Vulpecula, came in 1908. During the spring and summer of that year he concentrated on a part of the Milky Way in Cygnus and Aquila and some neighbouring regions. His method was initially to inspect the area with the naked eye and then with binoculars, numbering the stars as described above, although “After a while the sky became so familiar that the charts were but rarely consulted” (57). Until July 25 he saw nothing unusual, but on that particular night he found that one star seemed to be unusually faint. He observed it again on 26, 29 and 30 July and 2 August, but on these occasions the star appeared its normal brightness. Then on 3 August “the star was again faint, being one magnitude lower than on the previous evening”. Convinced that he had found a new variable he contacted Prof. H.H. Turner who reviewed Astbury’s data and then forwarded the discovery announcement to *Astronomische Nachrichten* (58) on 8 August.

The star, now known as RS Vul, was observed intensively by Astbury, Markwick and the Belgian variable star observer, Félix de Roy (1883-1942) (59). Astbury analysed their combined data and reported (60) a period of 4.4764 days and a visual brightness range of magnitude 6.9 to 8.0. The modern ephemeris lists the period as 4.47766 days (61).

Astbury’s RS Vul discovery paper was read on his behalf by Col. E.E. Markwick at the BAA meeting held in London on 25 November 1908. Markwick point out to the audience that “the discovery was made by an amateur armed only with an opera glass, and reflected great credit on Mr. Astbury for the care and discrimination shown in determining the fluctuations in light of this star" (62). Markwick had a further surprise up his sleeve for he went on to say that news had recently reached him that Astbury had discovered a further Algol type eclipsing binary, bringing his tally of discoveries to three. He was thus taking the opportunity of announcing this very latest discovery at the meeting, saying:

"the third discovery reflected the greatest credit, and even lustre, on the Variable Star Section in particular, and the British Astronomical Association in general."

The BAA President, H.P. Hollis, BA, FRAS (1858-1939) went on to comment that: (63)

"he used to think that the epithet 'patient' was most appropriately applied to the astronomer who searched for comets and found one in a year perhaps, but he was not sure now whether the variable star discoverer was not more patient still…..He asked members to congratulate Mr. Astbury, and said that he himself was pleased that the third discovery had happened in time to be announced at that Meeting."

The newly discovered eclipsing binary announced at the meeting is now known as Z Vul. As with the previous two discoveries, Astbury asked Prof. H.H. Turner to verify his observations, but it soon became apparent that the star in question had previously been reported (64) as variable by Albert S. Flint (1853-1923; Figure 12) of the Washburn Observatory, Wisconsin, USA in 1901. Flint had made his discovery whilst carrying out parallax measurements with the 12.2 cm meridian circle at Washburn and had actually used the star as one of his photometric standards until he found it varying (65). However, Flint had not determined a period. Astbury took Flint's times of eclipse minimum, made between 1898 and 1900, and combined them with his own from 1908 and determined a period of about 2.525 days (66). The following year, armed with further of times minimum determined from his own observations plus those of Flint and de Roy, he updated the period to 2.455 days and an amplitude between magnitude 7.3 and 8.5 (67). The modern value for the period is 2.45493 days (61). Thus Astbury had made an independent discovery of Z Vul and he and Flint should share the credit.

Astbury discovered a third eclipsing binary in 1911, TV Cas. He had monitored stars in the neighbourhood of β Cas during the month of September in 1908, 1909 and 1910. Then on the evening of 18 September 1911 he found one of the stars appeared to be very faint. He therefore followed it during the rest of September and into early October, allowing him to confirm its variability. Once again he shared his results with Prof. H.H. Turner and on 11 October announced his discovery in *Astronomische Nachrichten* (68). He also communicated news of his discovery to E.C. Pickering at Harvard College Observatory, who used the Observatory's extensive archive of plates to confirm the object's was variable and thus was able to provide some additional times of minima to update the ephemeris. Astbury sent his revised time of minimum ephemeris to Paul S. Yendell (1844-1918) of Dorchester, Massachusetts, USA. Yendell, a veteran of the American Civil War, was one of the most active American variable observers of the period, accumulating more than 30,000 observations (69), obliged by making further observations of TV Cas. Recent analysis of TV Cas's light curve suggests that it is a semi-detached binary system in which the Roche lobe-filling secondary is "spotty" (70), as shown in Figure 13.

Astbury's fifth and final confirmed variable was discovered in April 1913, with verification being provided by Prof H.H. Turner as had become the pattern (71). The star is now known as W UMi and is another eclipsing binary. This also transpired to be an independent discovery, shared with observers at the Royal Observatory, Greenwich (72). A recent light curve of W UMi obtained by VSS member Des Loughney using DSLR photometry (73) is shown in Figure 14. This clearly shows the deep primary and the shallow secondary eclipses in this semi-detached binary system. Astbury thought he had discovered another variable in the same field as W UMi (74), but this star, now known as NSV 7956, was soon shown not to be a variable.

In spite of Astbury's success in identifying new variables, there were three other cases where he mistakenly identified a star as variable. He thought he had found small variations in the star later named RT Vul, during his studies on the nearby RS Vul, which were discussed above. On 17 September 1908 and again on 27 October he recorded the star about half a magnitude fainter than normal. He reported his observations in a letter to *Astronomische Nachrichten* on 21 April 1909, suggesting it might be an Algol-type variable (75). Subsequent observations and modern data do not support this star being variable to any significant degree. A similar story unfolded in the case of VW Dra (reported in 1911) and NSV 13759 in Cepheus in 1910 (76). Astbury found the latter star to be about magnitude 9.5 (77). However, there is a further mystery in that the General Catalogue of Variable Stars (78) lists NSV 13759 as a magnitude 12.1 star (spectral type K5). It is rather surprising that Astbury could have observed such a faint star considering that his largest instrument was a 3¼ inch (8.3 cm) refractor (79). On the other hand there is a nearby $10^{th}$ magnitude star (TYC 4469-1254-1 = GSC 04469-01254 = 2MASS J21292045+7151275) which I consider a more likely candidate for Astbury's discovery, although it too is essentially of constant brightness. Northern Sky Variability Survey data for this star have a mean magnitude of 10.075±0.01 (80).

Astbury's variable star discoveries are listed in Table 1, along with a summary of their known properties. Thus between 1905 and 1913 he discovered 5 confirmed variables, 2 of which were independent discoveries shared with others. Of the 5 stars, one was a Cepheid (RT Aur) and the rest were Algol-type eclipsing binaries. Four further reported discoveries were subsequently found not to be variable at all, showing that whilst he was a very capable observer, he was not infallible.

## Arthur Eddington and the Cepheid variables

As we have seen, Astbury's first variable star discovery, RT Aur, was soon shown to be a Cepheid and it was this discovery that brought him to the attention of Prof. A.S. Eddington. Eddington was interested in understanding the internal constitution of stars and in 1916 he began to investigate possible physical explanations for the light variations of Cepheid variables, proposing that they were pulsating stars (81). His work on Cepheids later led him to generalise his ideas on the make-up of stars concluding that virtually all stars behaved as ideal gases. In 1926 he published his seminal book, *The Internal Constitution of the Stars,* which became one of the principal astrophysics textbooks of the age.

Although Eddington's work on Cepheids was not founded on the study of a specific star, after Astbury's death he did expound on his interest in RT Aur: (41)

“a discovery of more than ordinary interest fell to Mr. Astbury. His star RT Aur belongs to the class of Cepheids; and by reason of its brightness it is one of the half-dozen or so Cepheids which can be thoroughly investigated in all aspects. It is actually the third brightest of the ordinary Cepheids [now considered to be the eight brightest (82)]. Consequently it has figured conspicuously in all the studies of these stars which have furnished many surprising results; and astronomy owes a special debt to its discoverer…..In my own efforts to find out the constitution of the interior of a star I have had much guidance from RT Aurigae and its colleagues”.

Eddington’s work on Cepheid variability dealt a death blow to one of the more popular alternative explanations of the time: that they were eclipsing binary systems. In fact in 1909 one researcher at the Lick Observatory had even gone so far as to publish the orbital elements of RT Aur and another bright Cepheid, Y Sgr (83). However, even today RT Aur is still studied by astrophysicists and it continues slowly to reveal its secrets. A 2007 paper (84) by Prof. David Turner of Saint Mary's University in Halifax, Canada, and his co-workers shows that its nominal 3.7 day pulsation period is actually increasing at a rate of 0.082±0.012 seconds/year. They looked at the difference between the observed (O) and calculated (C) times of maxima very closely and they found superposed on the star's O-C variations a subtle sinusoidal trend. They could not attribute this effect to random fluctuations in the pulsation period. Rather, they suggest it arises from the effects of light travel time in a binary system. Their derived orbital period for the system is P = 26,429±89 days (72.36±0.24 years). So it appears that RT Aur may be a binary system after all, with one component a Cepheid variable. It’s also interesting to note that the researchers used amateur observations of the star going back to 1973 in their analysis (Figure 15), once again highlighting the value of long-term amateur data contributing to the development of variable star science in particular and astrophysics in general. In addition, it is interesting to note that the authors made used recent CCD photometry obtained with typical amateur instrumentation. The importance of continued monitoring of Cepheid variables by amateur astronomers in an era when professional observations of such stars are declining (85) is illustrated clearly by the case of RT Aur.

**The Wallingford school memorial**

As we saw earlier, Astbury retired as Headmaster of Wallingford Council School for Boys in 1920 on the grounds of ill health, having served the school community of Wallingford for 36 years. He died on 28 September 1922 aged 64, “his funeral being attended by representatives of education and science. Many of his old pupils can testify to the great interest Astbury took in his boys, and the kindness and help he showed them both at school and in after years” (41).

The idea of erecting a tablet in his memory originated from H.H. Turner, who as we have seen was Astbury’s mentor for many years. E.E. Markwick, who also had a long association with Astbury through the BAA VSS (86), immediately threw his weight and enthusiasm behind the project. Knowing Eddington’s interest in Cepheids his support was also sought and readily given. A.N. Brown, VSS Secretary, collected the donations from supporters. The four then placed the notice in the BAA *Journal* which was cited in the Introduction, with similar appeals for support appearing in *The Observatory* (87) and the *English Mechanic* (88). Twenty-two subscribers supported the appeal fund.

The resulting brass memorial tablet, shown in Figure 16, was erected at Wallingford Council School for Boys and was unveiled by Sir Frank Dyson (1868 – 1939) on 29 May 1924. Dyson was Astronomer Royal and had a long association with the BAA, including serving as President from 1916 to 1918. He would therefore have been familiar with Astbury's work. Dyson gave a talk during which he paid tribute to Astbury's variable star discoveries and explained in simple terms the pulsation theory of Cepheid variables. Although Eddington was not able to present in person, he did send a letter which was read on his behalf by A.N. Brown. Eddington praised Astbury's discovery of RT Aur and described how the study of Cepheid variables, such as RT Aur, had led him to develop his ideas on the internal constitution of the stars, noting: (41)

"It is a real pleasure to associate myself with this tribute to one who laboured well for the advancement of astronomy. The study of variable stars, to which he devoted his leisure, is becoming more and more important, and is fundamental in much of the recent progress of our knowledge. The army of amateur observers who examine the stars night by night, as he did, to obtain the foundation-data on which so much is to be built, have little idea of achieving individual fame for themselves; they love their work, and they know that in due time progress will come from the united efforts of the whole band"

H.H. Turner went on to describe how Astbury had discovered RT Aur and conveyed a message of appreciation and sympathy from the Herschel family (89). Col. Markwick also gave a speech highlighting the work of the BAA VSS and Astbury's contribution to the field of variable star research. He warmly described the many letters they had exchanged over the years, especially around the time of Astbury's discovery of RS Vul. A tribute was also read from Prof. R.A. Sampson (1866-1939), Astronomer Royal for Scotland, who said: (41)

"Men like Astbury appear to me some of the best on earth. Their single-hearted love of science may very easily impress others more, and reproduce itself better than does the work of a professional"

Other speakers at the ceremony honoured Astbury's contribution to education, including Alderman H.W. Wells, Chairman of the School Managers, and Mr. Anderson, Secretary to the Berkshire Education Committee.

The afternoon ended with a tea laid out in Wallingford Town Hall, hosted by the Mayor, during which one of Astbury's daughters thanked Turner, on her mother's behalf, for organising the tribute. Turner had "been a most kind and generous friend of the Astbury family" (90).

**Astbury's place amongst the stars**

It is sometimes thought that links between professional and amateur astronomers are a modern phenomenon (91). However, as Bob Marriott notes: (92) "A century and more ago it was natural for astronomers to work together, whatever their position or status, and many professional astronomers – including H.H. Turner – were members of the BAA". Thus the association between Astbury and Turner was not unique. Initially Turner was in a supporting role, verifying his observations and ensuring they were communicated properly and published in the literature, where they could be brought to the attention of other astronomers. Then as Astbury gained more experience he became confident enough to make his own announcements. Later, of course, Astbury's discoveries, RT Aur in particular, were taken up by professionals and noticed by the likes of Eddington. It is against this background of mutual respect that led some of the country's most famous professional astronomers to erect the memorial to Astbury.

During the course of this research, I noticed that none of Astbury's 5 confirmed variable star discoveries were credited to him in the AAVSO Variable Star Index (VSX). I am delighted to report that with the help of one of the VSX managers, Patrick Wils, this is in the process of being corrected. The entry for RT Aur now proudly cites the name of T.H. Astbury as discoverer (Figure 17). A fitting tribute for a long-forgotten amateur astronomer.

**Acknowledgements**

I am most grateful for the assistance I have received from a great many people whilst preparing this paper. Special thanks are due to Tracey Clark & Emma Anderson for permission to reproduce several photographs from their book *St John's County Primary School Wallingford: Celebrating One Hundred Years of a Oxfordshire Market Town School*, some of which were first published in *Wallingford County Boys School 1910-1971* by Arthur Dean. Tracey was also kind enough to provide all sorts of background information about the School. Hazel McGee kindly researched Astbury's Census returns. Patrick Wils pointed me in the direction of NSVS data on TYC 4469-1254-1, the star that I propose as Astbury's "constant" variable star in Cepheus, NSV 13759. He also updated the AAVSO Variable Star Index, crediting Astbury's variable star discoveries. Arne Henden, AAVSO Director, gave permission to use the RT Aur listing from the AAVSO International Variable Star Index. Richard Baum, who has provided much encouragement in my forays into astronomical history, led me to appreciate the significance of Astbury's observation of the occultation of Saturn and suggested that this warranted further study, the results of which we will report separately in a joint paper. Des Loughney, BAA VSS Eclipsing Binary Secretary, undertook a programme of DSLR photometry to produce the phase diagram of W UMi, one of Astbury's discoveries. Prof. David Turner, Saint Mary's University, Halifax, Canada, gave permission to use his figure from reference (84) containing light curves of RT Aur. Mike Frost obtained copies of some of Astbury's published notes. Jill Barlow searched the Cheltenham College Archives for records related to Astbury and was able to rule out his having attended the college, in spite of the statement by Col. E.E. Markwick that he had. Hugh Darlington supplied the previously unpublished photograph of Charles Lewis Brook.

This research made use of the NASA/Smithsonian Astrophysics Data System, the AAVSO Variable Star Index, ROTSE data accessed via the Northern Sky Variability Survey operated through Los Alamos National Laboratory, and SIMBAD and Vizier, operated through the Centre de Données Astronomiques (Strasbourg, France). I also made extensive use of scanned back numbers of the *Journal*, a truly wonderful resource, which is largely thanks to the efforts of Sheridan Williams, the existence of which saved me several trips to libraries to consult the printed word.

Finally I thank the referees, Mike Frost and Bob Marriott for their helpful suggestions that have improved the paper.

**Address**

"Pemberton", School Lane, Bunbury, Tarporley, Cheshire, CW6 9NR, UK
[bunburyobservatory@hotmail.com]

30. The collapse of the Meteor Section in 1905 when the Director, W.E. Beasley, died suddenly is described in the BAA memoir on the First Fifty Years. Between 1906 and 1911 there was an interregnum. Astbury is cited in the Memoir as an active member of the Section in the years leading up to the collapse

31. An example was a meteor on 16 April 1906, which was observed by Astbury at Wallingford as being brighter than Venus and was also seen by Denning "and four friends on Horfield Common in Bristol", see: Denning W.F., English Mechanic, May 4 1906, item 319.

32. Astbury wrote a series of 3 letters on the zodiacal light observed from Wallingford in February and March 1899: Astbury T.H., JBAA, 9, 328-330 (1899).

33. He observed an aurora on 12 October 1903 and commented that a magnetic needle showed distinct oscillations at the same time: Astbury T.H., JBAA, 14, 28 (1903). There was a magnificent auroral display on 15 November 1905 which he observed: Astbury T.H., JBAA, 14, 28 (1903). The same edition of the BAA Journal carries reports by other observers across the country.

34. He wrote occasionally about his comet observations in the English Mechanic. He also described how he saw the nucleus of Comet 1900 II almost occult a field star: Astbury T.H., JBAA, 11, 281 (1901). He also describes his observations of Halley's Comet in February 1910 (Astbury T.H., English Mechanic, 2431, 4 Feb 1910).

35. Astbury T.H., JBAA, 11, 35-36 (1900) describes his observation of the occultation with his 3¼ inch Wray under almost perfect conditions.

36. Astbury T.H., JBAA, 15, 40 (1904).

37. Astbury's observation of the Saturnian prominence is discussed in more detail in a separate note by the author and Richard Baum: Shears J. & Baum R., JBAA, accepted for publication (2012).

38. Col. E.E. Markwick presented a report on the VSS observations of the nova to the BAA meeting of 24 April 1901 in which he drew attention to Astbury's observations during the period 18-31 March when the brightness was fluctuating. Astbury had noted that as it brightened, so it became redder.

39. Markwick E.E., English Mechanic, 2056, 40, item 43 (1904).

40. Brook succeeded Markwick as VSS Director after Markwick resigned in December 1909. Brook's life is described in Shears J., JBAA, 112, 17-30 (2012).

41. "The late T.H. Astbury: An Appreciation" appears as: Markwick E.E., JBAA, 34, 327-331 (1924).

42. The 1900-1904 and 1905-1909 memoirs are Markwick E.E., Mem. Brit. Astron. Assoc., 15 (1906) and Markwick E.E., Mem. Brit. Astron. Assoc., 18 (1912) respectively.

43. Brook C.L., Mem. Brit. Astron. Assoc., 22 (1918).

44. Toone J., JBAA, 120, 135-151 (2010).

93. Astbury T.H., JBAA, 18, 132 (1907).

| Star | Preliminary designation | Year | RA Dec (J2000) | Type | Period | Range (mag.) | Remarks |
|---|---|---|---|---|---|---|---|
| **Unique discoveries** | | | | | | | |
| RT Aur | 47.1905 | 1905 | 06 28 34.09 +30 29 34.9 | Cepheid | 3.728115 | 5 - 5.82 V | |
| RS Vul | 16.1908 | 1908 | 19 17 39.99 +22 26 28.4 | EA/SD | 4.4776635 | 6.79 - 7.83 V | |
| TV Cas | 45.1911 | 1911 | 00 19 18.74 +59 08 20.6 | EA/SD | 1.8125956 | 7.22 - 8.22 V | |
| **Independent discoveries** | | | | | | | |
| Z Vul | 26.1900 | 1908 | 19 21 39.10 +25 34 29.5 | EA/SD | 2.454934 | 7.25 - 8.9 V | Independently discovered by Flint in 1900 (64) |
| W UMi | 12.1913 | 1913 | 16 08 27.27 +86 11 59.6 | EA/SD | 1.7011576 | 8.51 - 9.59 V | Independent discovery at RGO (72) |
| **Unconfirmed variables** | | | | | | | |
| RT Vul | 13.1909 | 1909 | 19 11 28.82 +22 22 59.3 | CST | -- | 7.5 p | |
| VW Dra | 1.1911 | 1911 | 17 16 29.40 +60 40 14.2 | CST | -- | 6.32 V | |
| NSV 13759 | 39.1910 | 1910 | 21 29 19.3 +71 56 30.3 | CST | - | - | Astbury identified it as BD +71°1070 ~9.5 v. Possibly TYC 4469-1254-1 at RA 21 29 20.5 Dec +71 51 27.5 (see text) |
| NSV 7956 | 13.1913 | 1913 | 16 29 23.32 +86 26 00.9 | CST | - | - | In the same field as W UMi and reported by Astbury at the same time |

Table 1: Astbury’s variable star discoveries

Preliminary designation is from *Astronomische Nachrichten*. An EA/SD star is an Algol-type eclipsing binary. CST = constant. Data from the AAVSO Variable Star Index (VSX) except for AN 39.1910 where the data are from the General Catalogue of Variable Stars

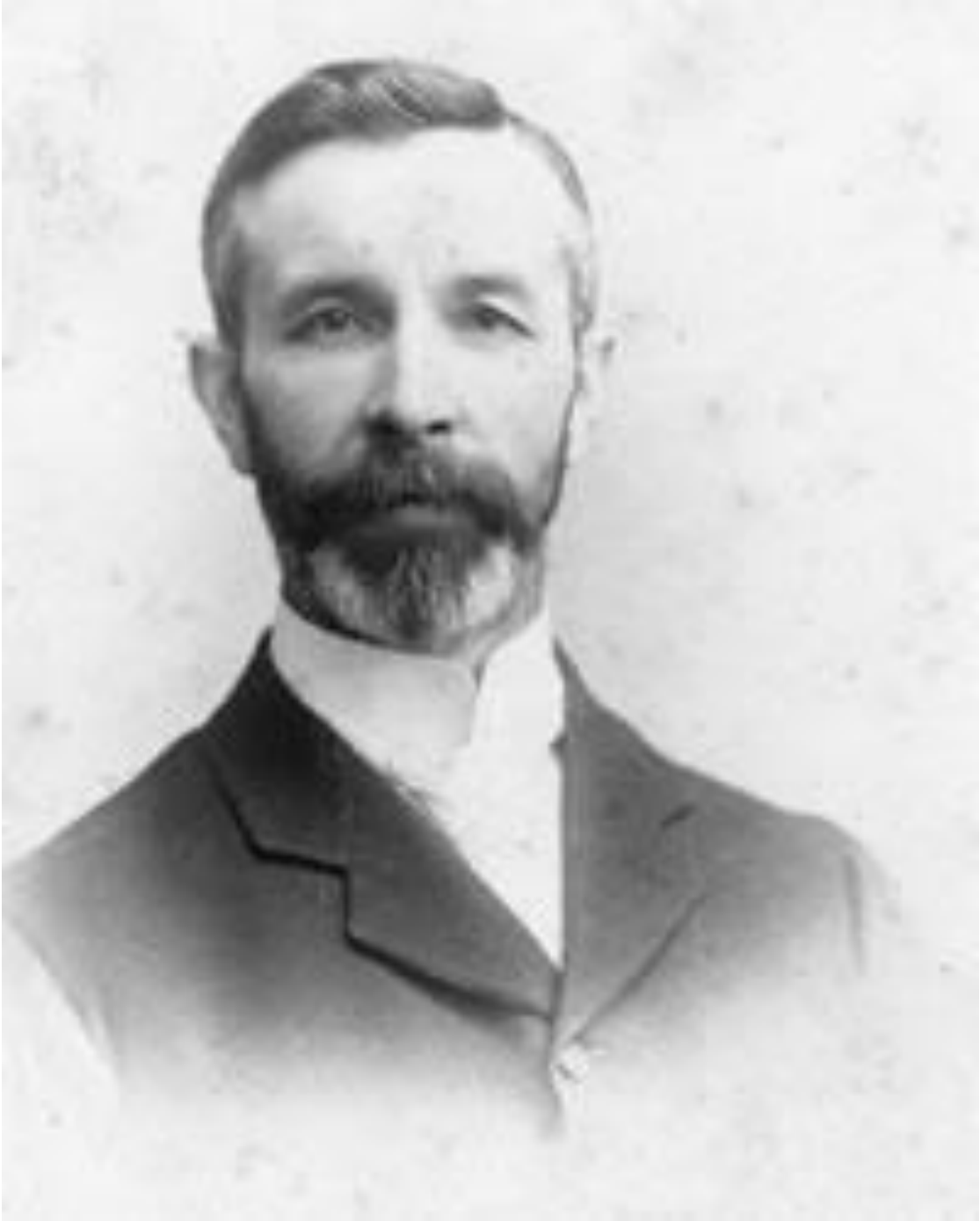

Figure 1: Thomas Hinsley Astbury (1858-1922). From reference (15)

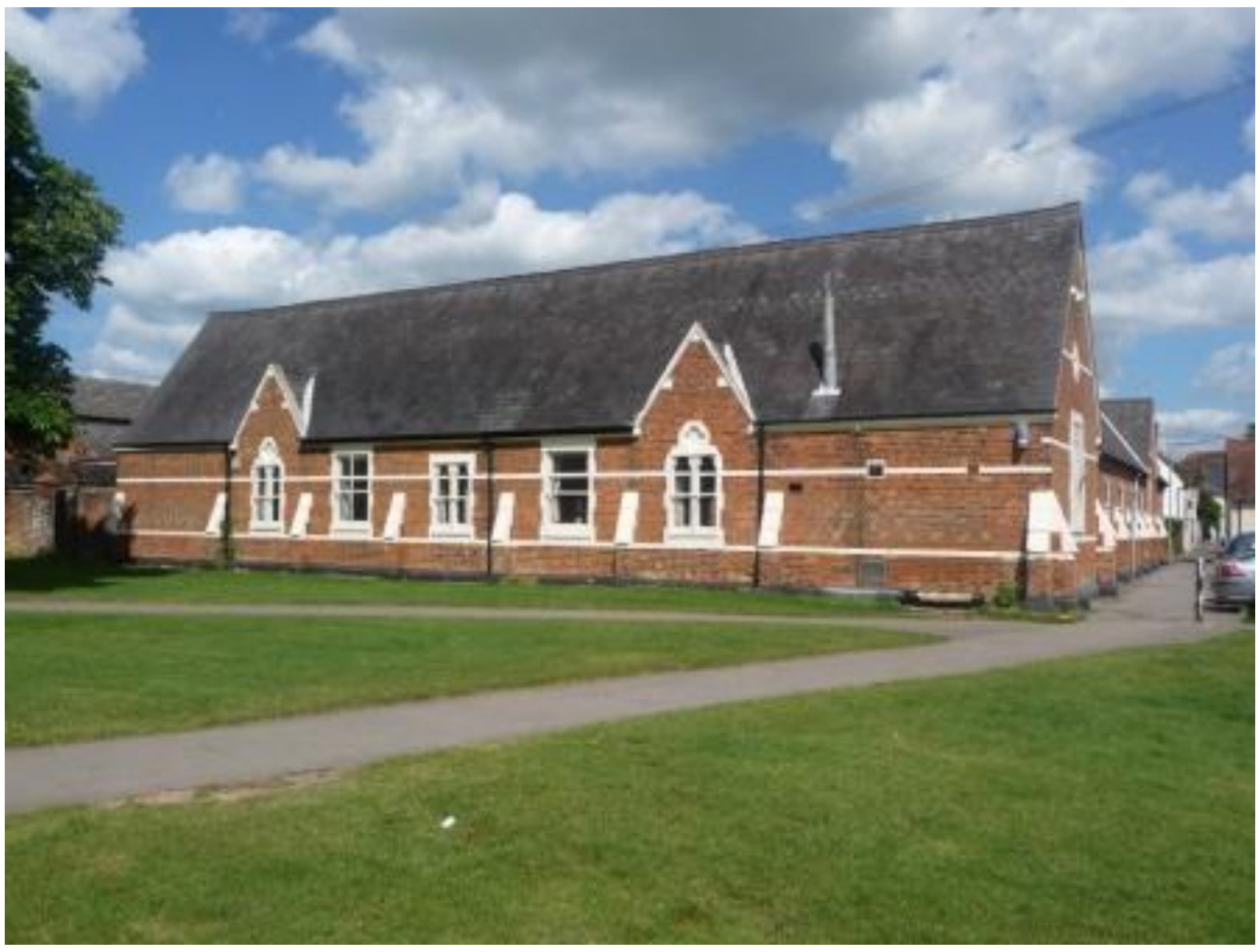

Figure 2: Kinecroft School

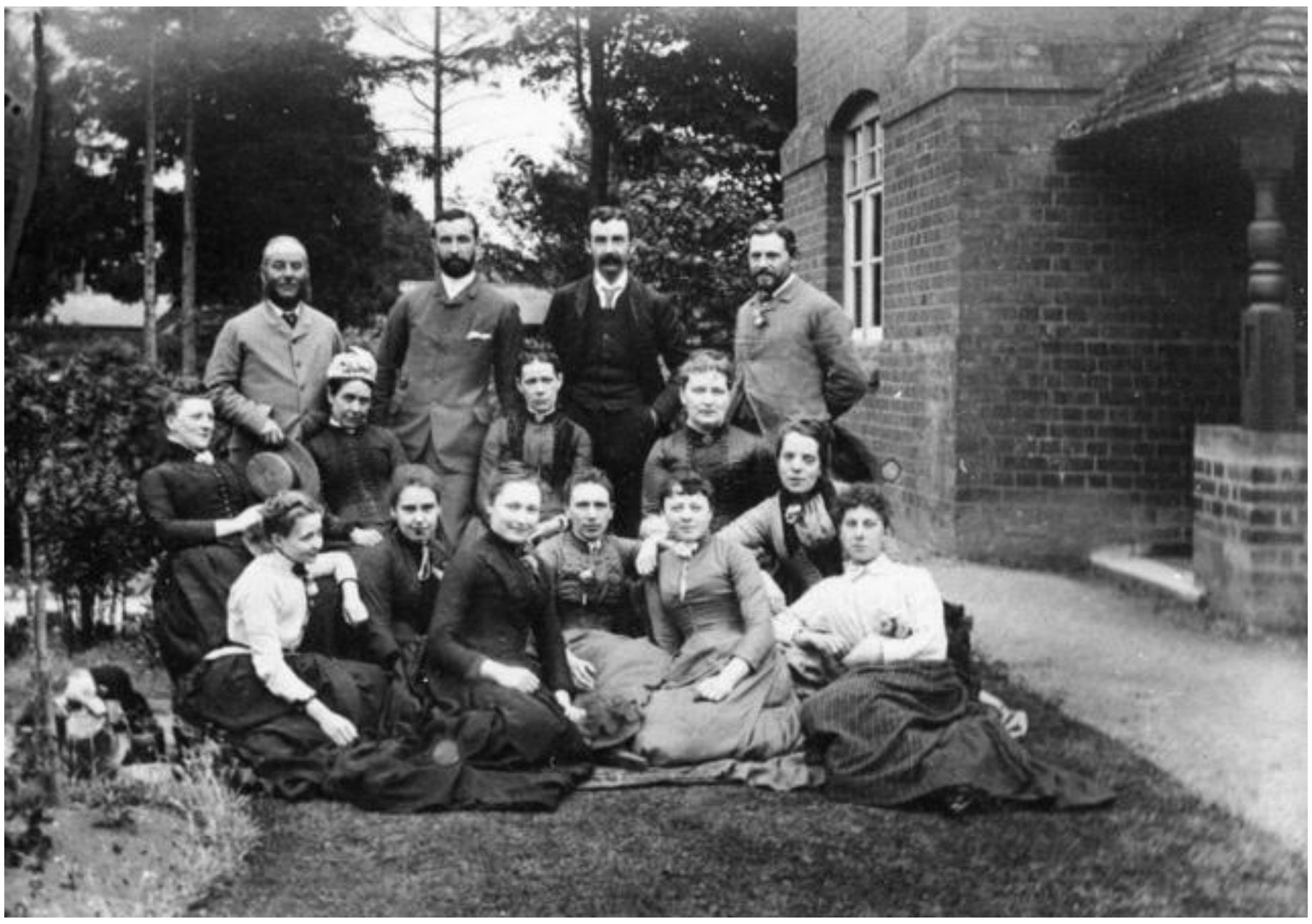

Figure 3: T.H. Astbury (back row, second from left) and other schoolteachers at Brightwell Park, near Wallingford, ca. 1900. From reference (15)

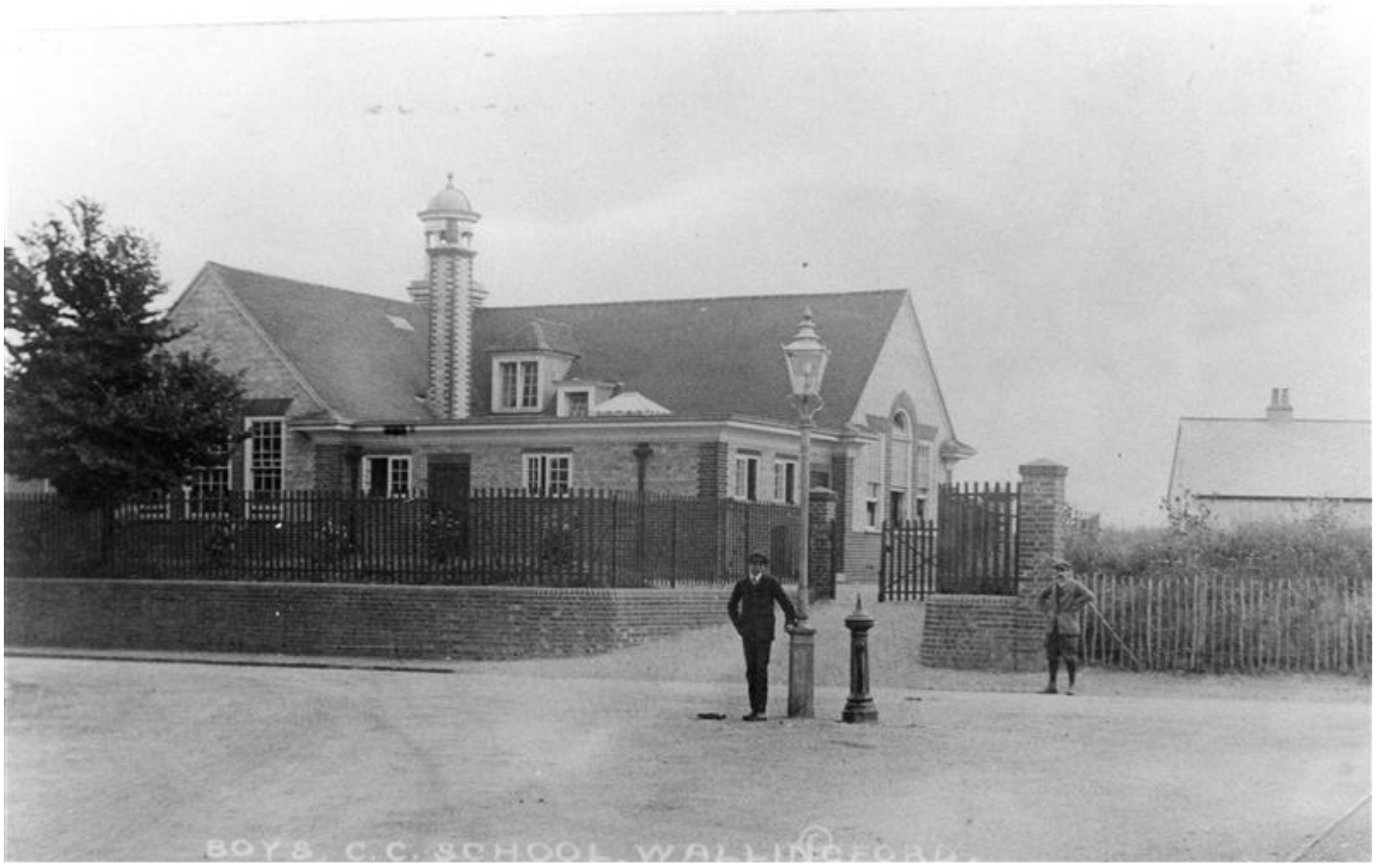

Figure 4: Wallingford Council School for Boys, ca. 1910. From reference (15)

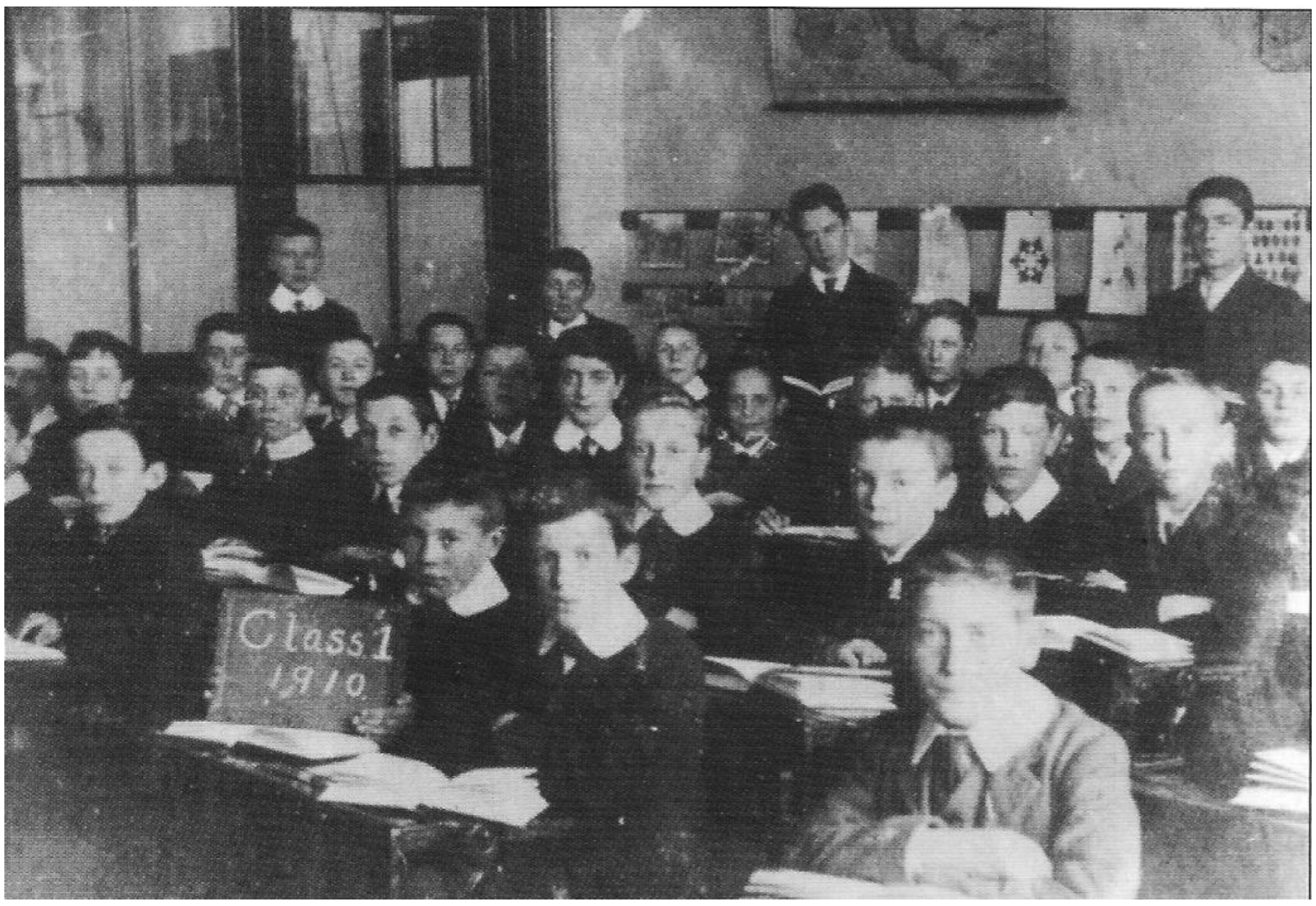

Figure 5: Class 1 at Wallingford Council School in 1910. From reference (15)

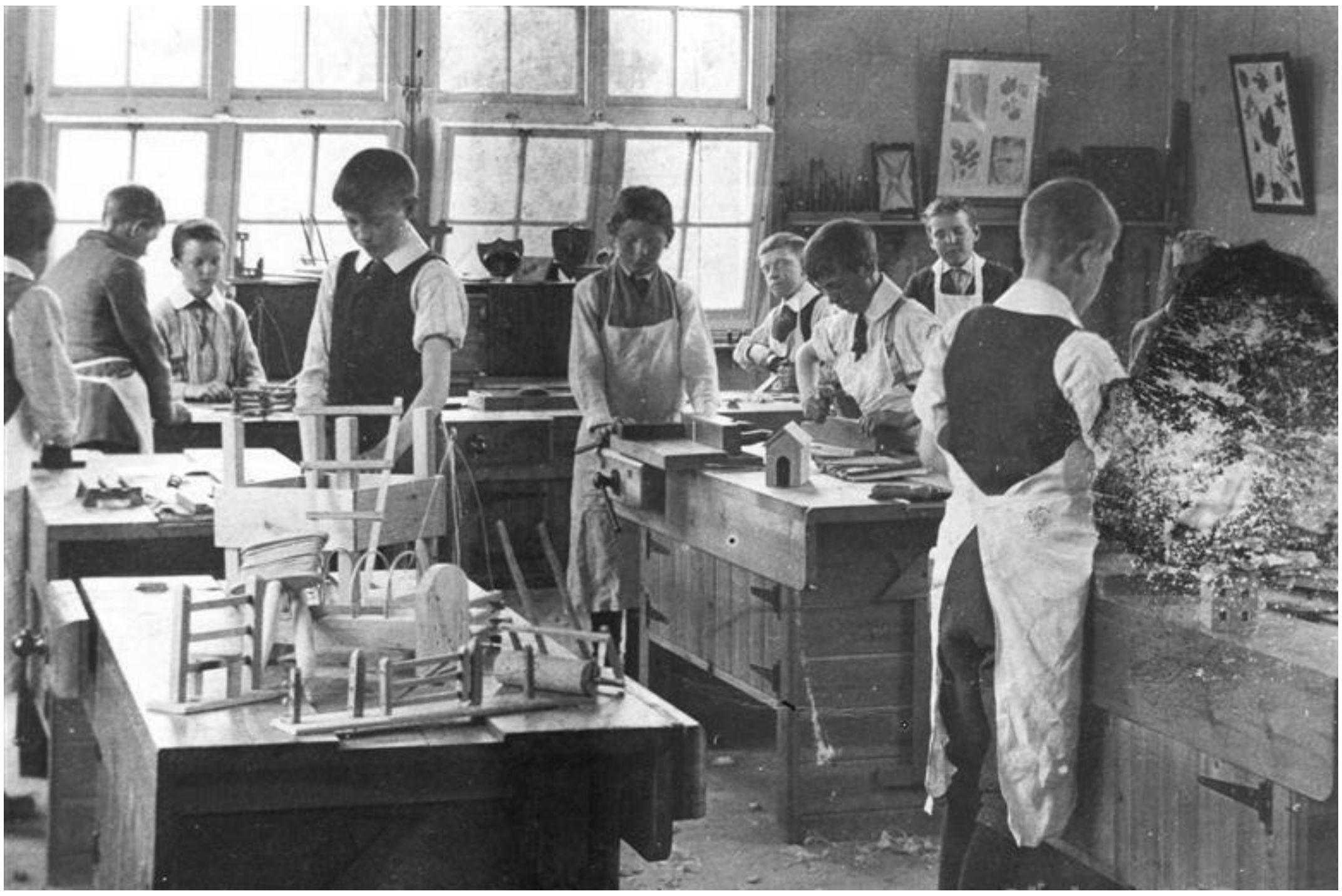

Figure 6: A woodwork lesson in progress at Wallingford Council School for Boys, ca. 1912, shortly after the opening of the Cooking and Manual Centre. From reference

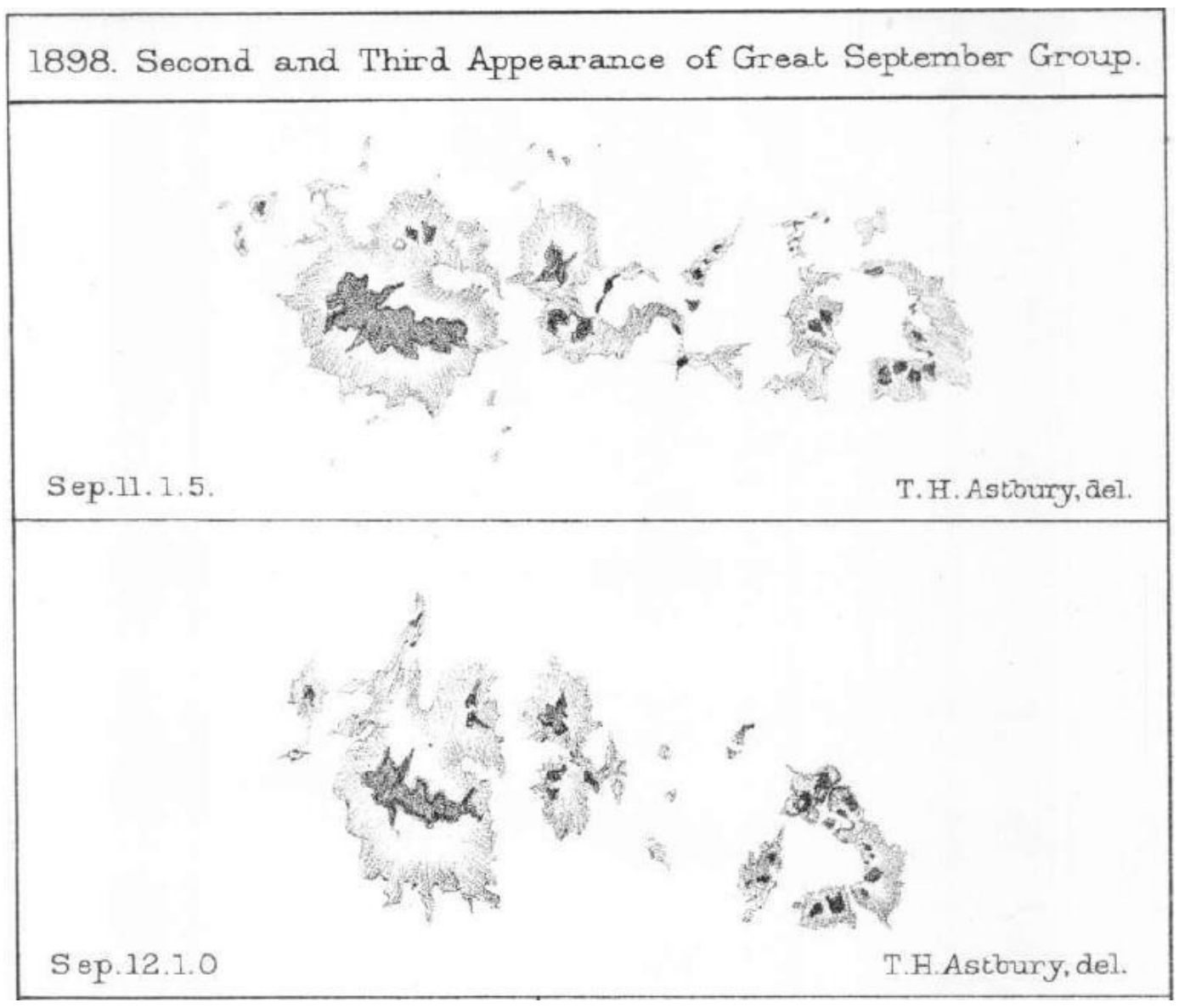

Figure 7: Drawings by Astbury of the Great Sunspot Group of September 1898

OBSERVATIONS OF (CH. 1226) NOVA PERSEI.

| Date. | Class. | Mag. | Observer. |
|---|---|---|---|
| 1903. | | | |
| March 3 ........ | 1 | 9 55 | Markwick |
| ,, 16 ........ | 1 | 9·42 | ,, |
| April 12 ........ | 1 | 10·1 | Brook |
| Aug. 25 ........ | 1 | 10·00 | Brook |
| ,, 25 ........ | 1 | 9 95 | Markwick |
| Sept. 23 ........ | 2 | 9·9 | Astbury |
| ,, 25 ........ | 2 | 10 0 | ,, |
| Oct. 23 ........ | 1 | 10·0 | Brook |
| Nov. 7 ........ | 2 | < 9·8 | Markwick |
| ,, 14 ........ | 1 | 9 9 | ,, |
| ,, 25 ........ | 2 | 10 1 | Brook |
| 1904. | | | |
| Jan. 16 ........ | 1 | 10 25 | Markwick |
| ,, 19 ........ | 1 | 10·0 | ,, |
| Feb. 6 ........ | 1 | 10·3 | ,, |
| ,, 13 ........ | 1 | 10·3 | ,, |
| ,, 14 ........ | 1 | 10 3 | ,, |
| ,, 15 ........ | 1 | 10 45 | ,, |
| ,, 18 ........ | 1 | 10·5 ± | ,, |
| March 8 ........ | 3 | { much < 9 | Astbury |
| ,, 10 ........ | 2 | 10·2 | ,, |
| ,, 10 ........ | 1 | 9 87 | Markwick |
| April 9 ........ | 1 | 10·0 | ,, |
| ,, 12 ........ | 1 | 9·8 | ,, |

E. E. Markwick, Col.
Salisbury, August, 1904.

Figure 8: Observations of Nova Persei 1901 made during 1903 and 1904, reported to E.E. Markwick and published in the *English Mechanic* (39)

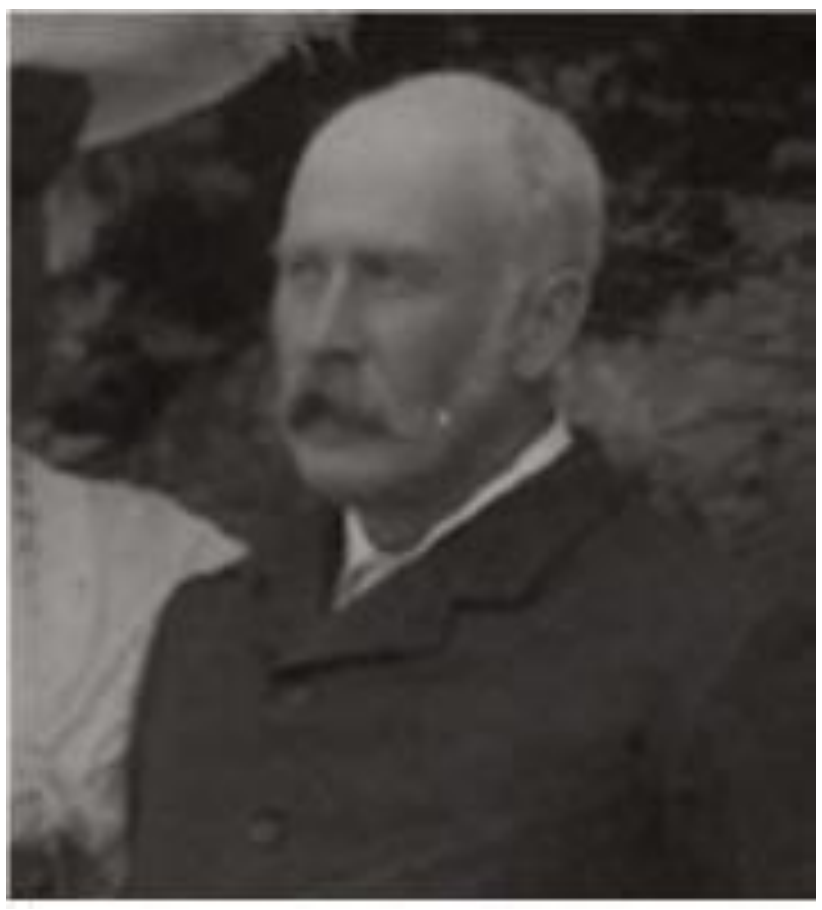

Figure 9: Charles Lewis Brook, MA, FRAS, FRMetS (1855-1939) photographed in 1909. Photograph courtesy of Hugh Darlington

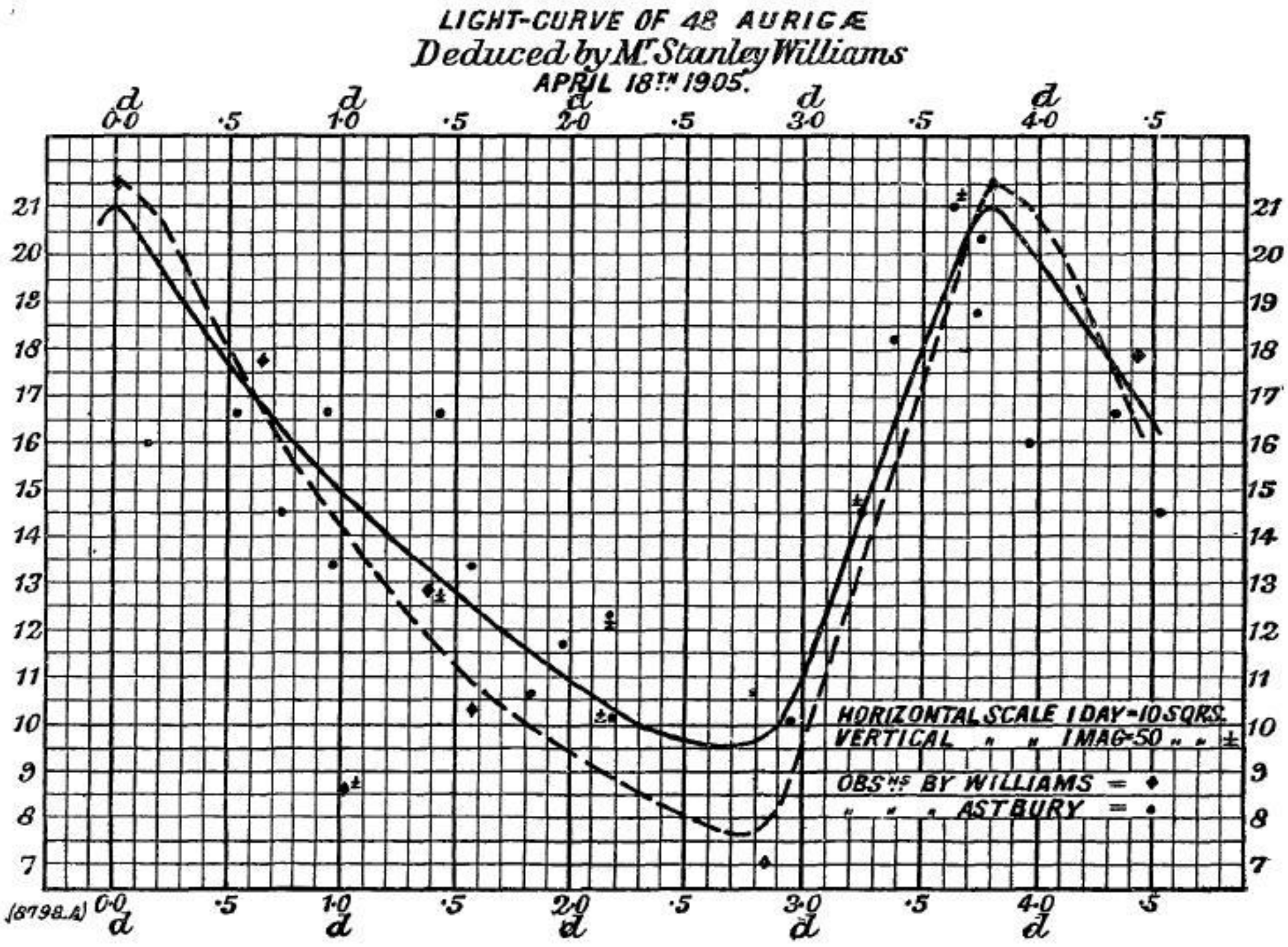

Figure 10: Light curve of RT Aur drawn by Arthur Stanley Williams, based on his own and Astbury's observations between 18 March and 12 April 1905. Stanley Williams plotted the data based on his assumed period of 3.8 days. Horizontal axis: one day = 10 squares. Vertical axis: 1 magnitude = 50 squares. From reference (52)

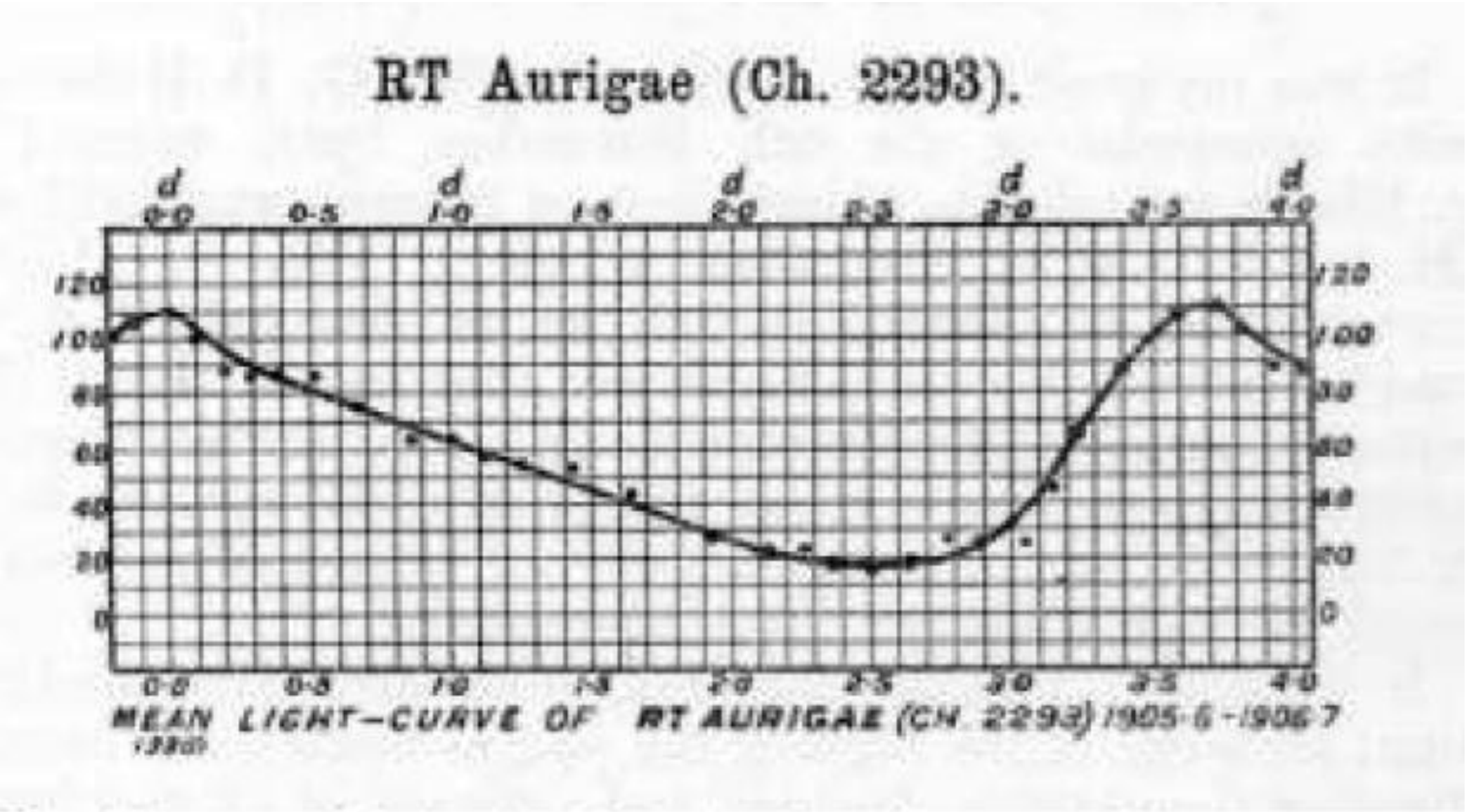

Figure 11: Astbury's revised light curve of RT Aur containing data from 1905 to 1907. From reference (93)

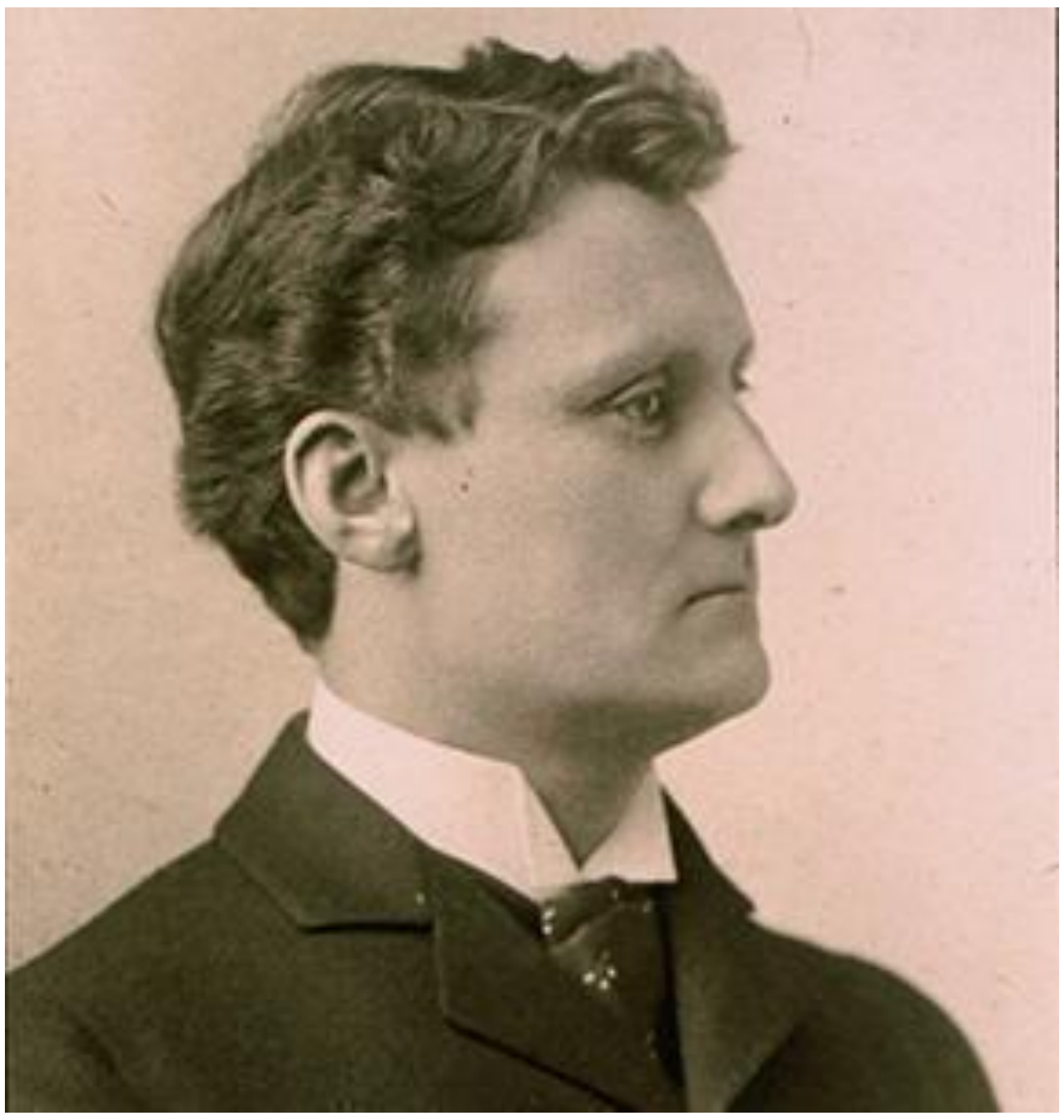

Figure 12: Albert Stowell Flint (1853-1923) of the Washburn Observatory

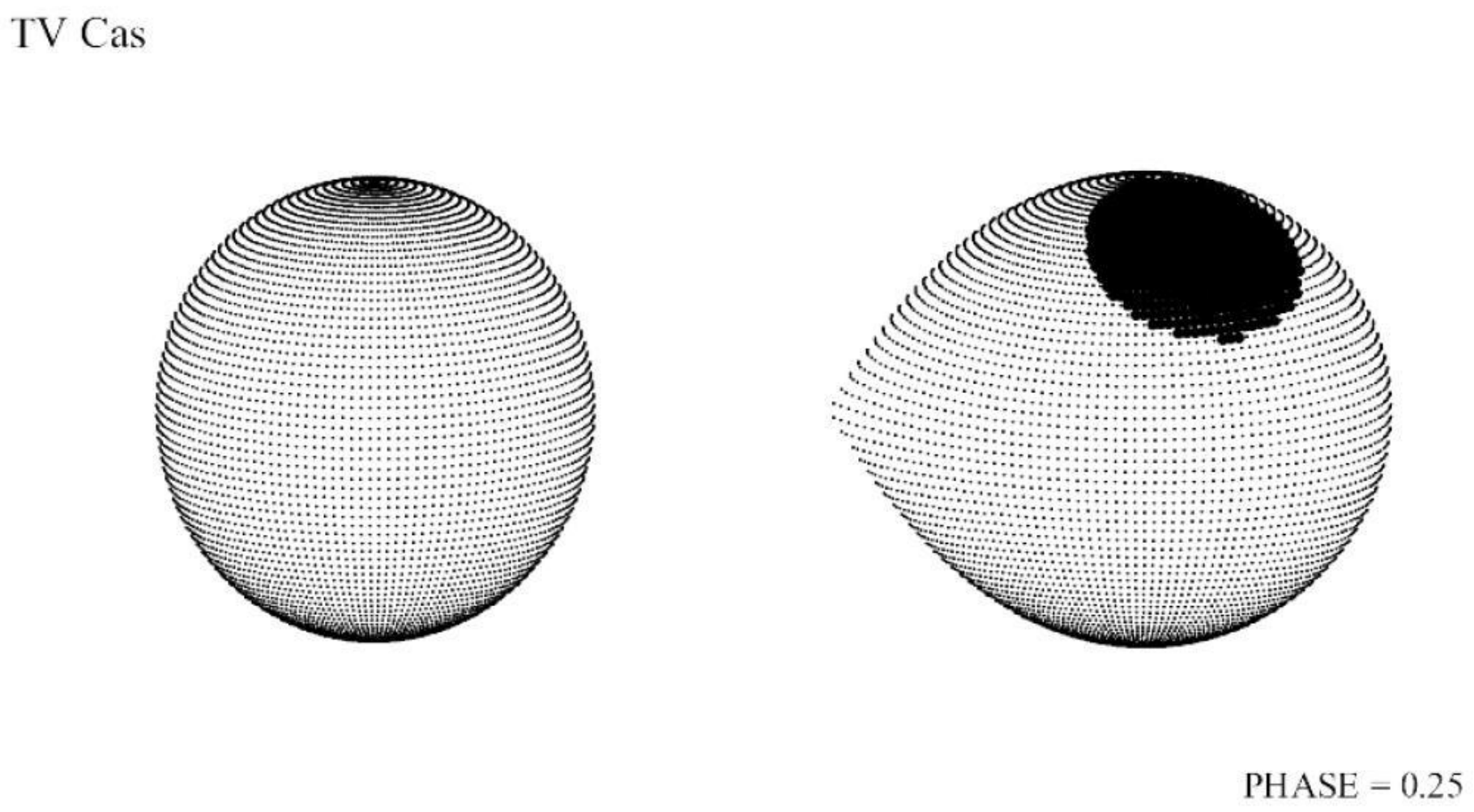

Figure 13: View of the TV Cas system at orbital phase 0.25, showing the proposed "star spot" on the secondary. From reference (70)

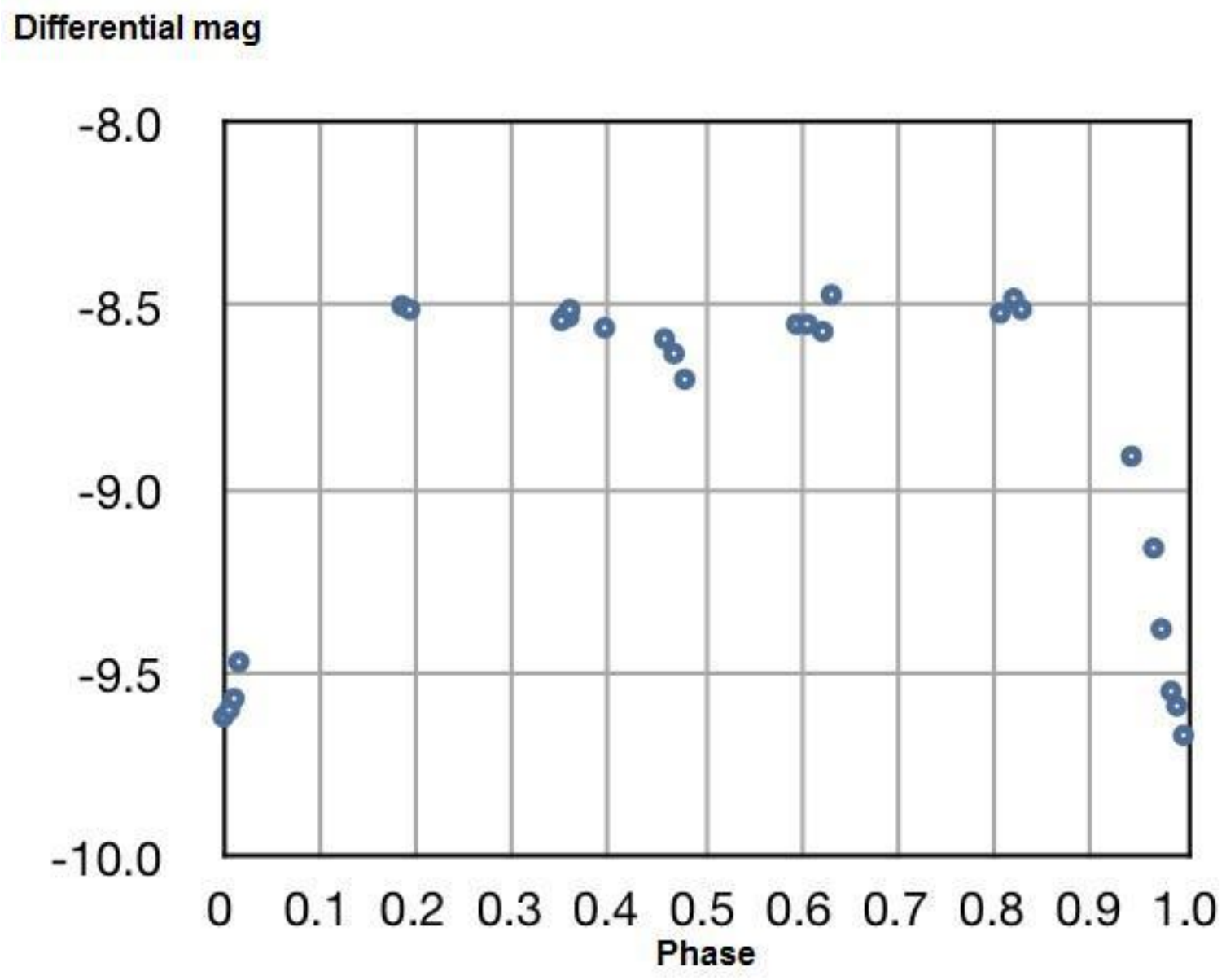

Figure 14: Phase diagram of W UMi, obtained by Des Loughney based on photometry with a DSLR camera (note this diagram will be updated with further data points)

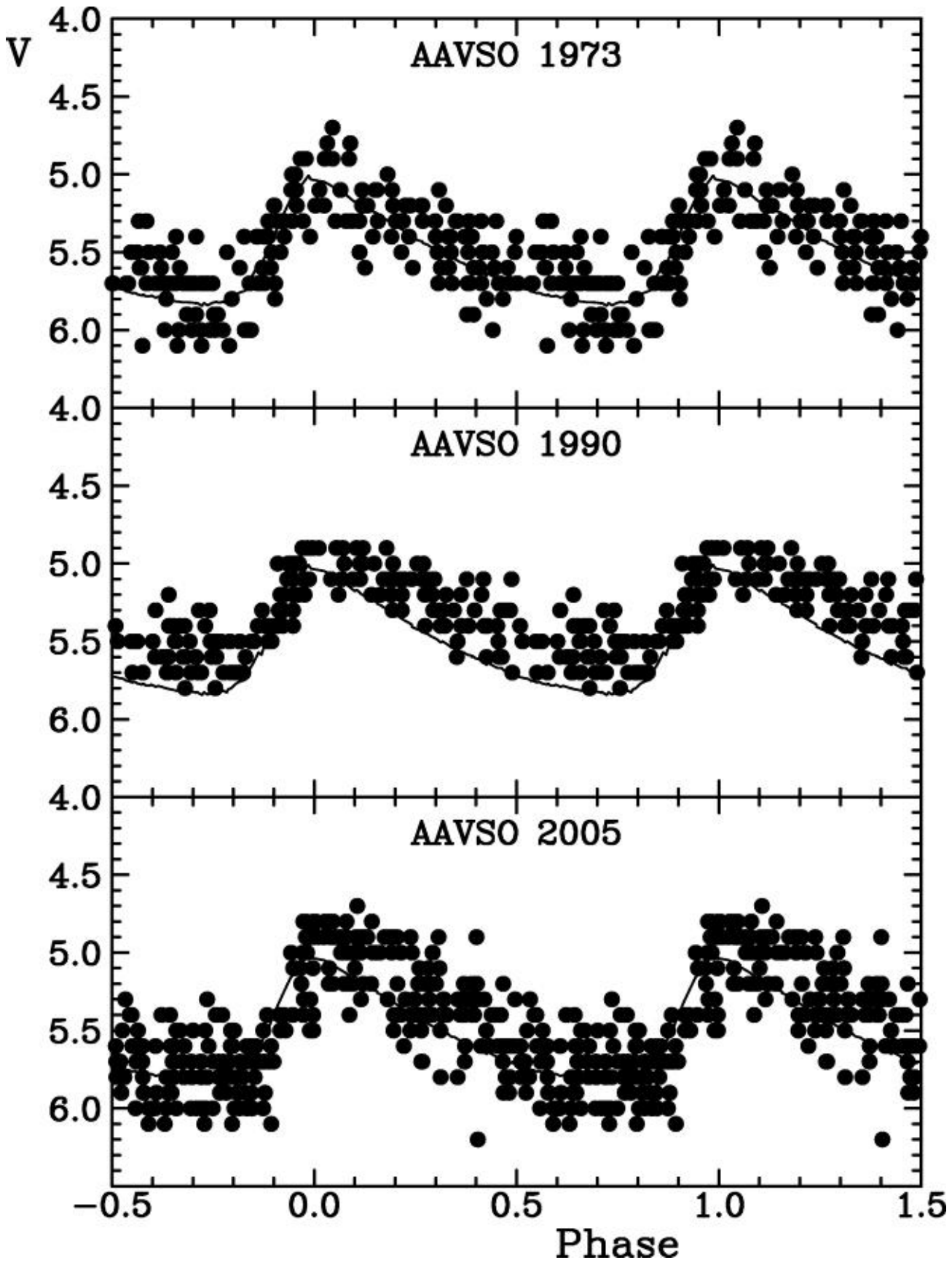

Figure 15: Sample of yearly phase diagrams for RT Aur using data from the AAVSO International Database. Figure from Turner D.G. *et al.* (reference (84)

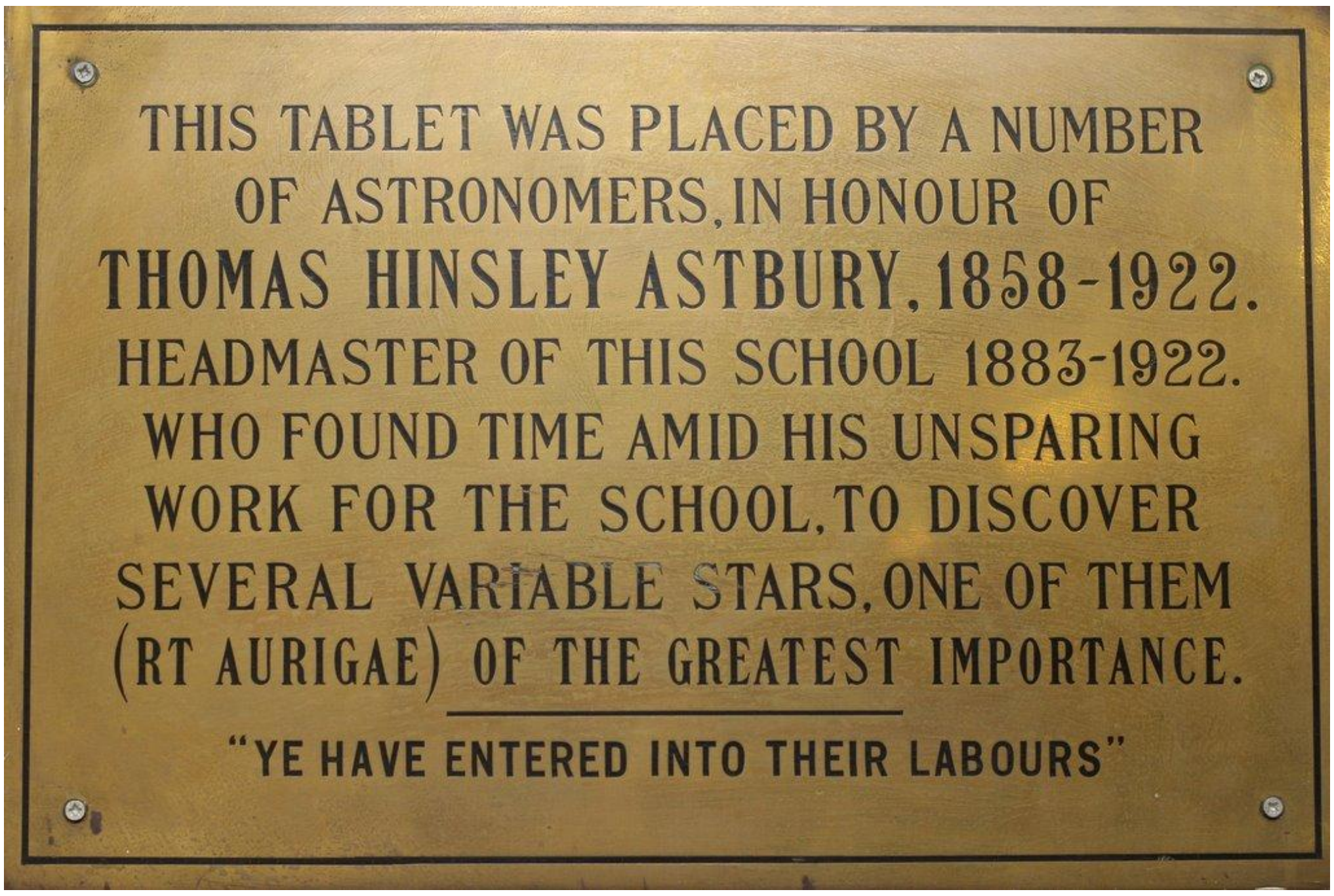

Figure 16: Memorial tablet to Astbury.

Note that there is an error in the dates during which he was headmaster: he retired in 1920

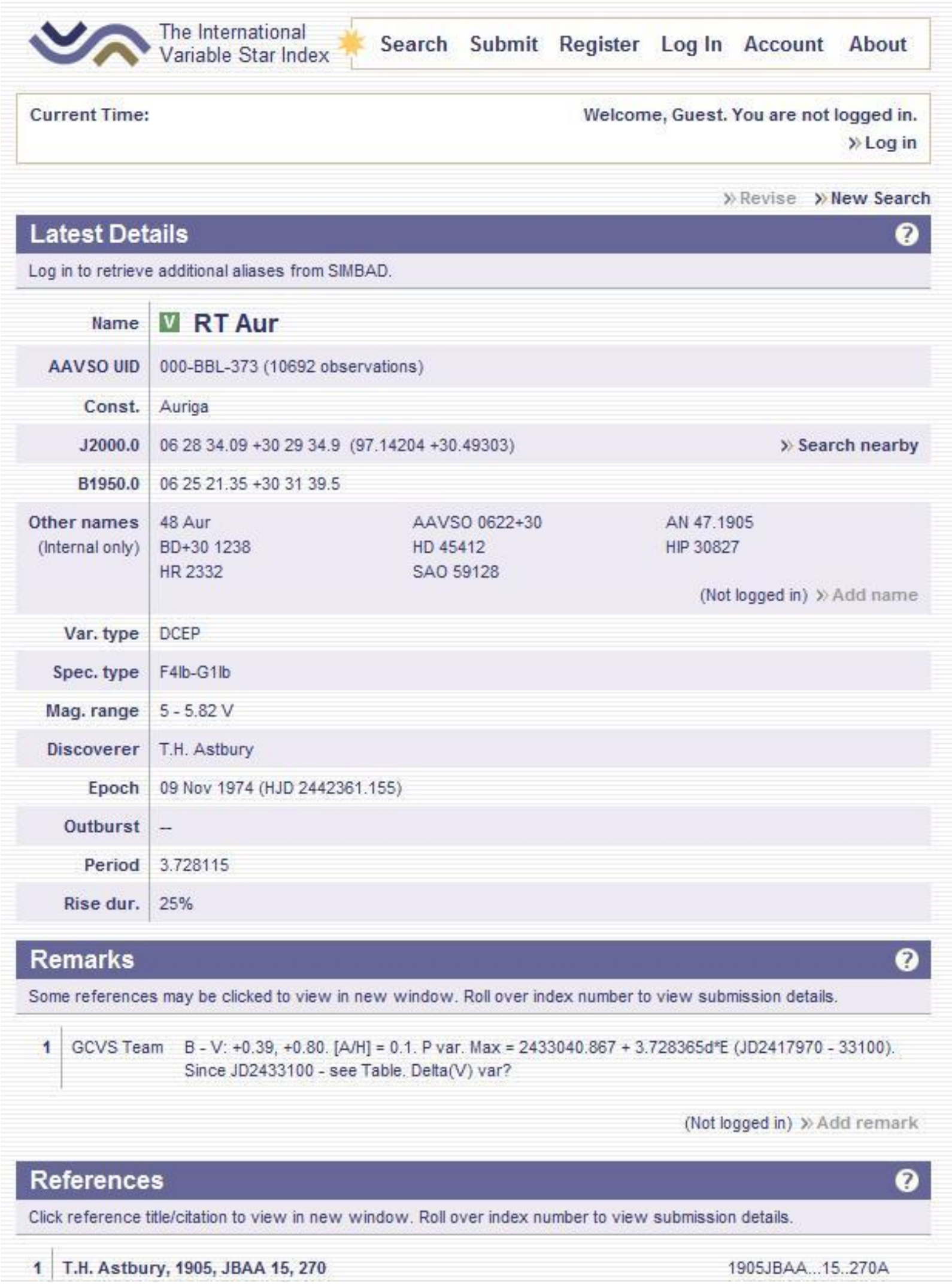

The International Variable Star Index | Search | Submit | Register | Log In | Account | About

Current Time: | Welcome, Guest. You are not logged in. » Log in

» Revise » New Search

**Latest Details**

Log in to retrieve additional aliases from SIMBAD.

| | |
|---|---|
| Name | V RT Aur |
| AAVSO UID | 000-BBL-373 (10692 observations) |
| Const. | Auriga |
| J2000.0 | 06 28 34.09 +30 29 34.9 (97.14204 +30.49303) » Search nearby |
| B1950.0 | 06 25 21.35 +30 31 39.5 |
| Other names (Internal only) | 48 Aur, BD+30 1238, HR 2332, AAVSO 0622+30, HD 45412, SAO 59128, AN 47.1905, HIP 30827 (Not logged in) » Add name |
| Var. type | DCEP |
| Spec. type | F4Ib-G1Ib |
| Mag. range | 5 - 5.82 V |
| Discoverer | T.H. Astbury |
| Epoch | 09 Nov 1974 (HJD 2442361.155) |
| Outburst | -- |
| Period | 3.728115 |
| Rise dur. | 25% |

**Remarks**

Some references may be clicked to view in new window. Roll over index number to view submission details.

1 GCVS Team B - V: +0.39, +0.80. [A/H] = 0.1. P var. Max = 2433040.867 + 3.728365d*E (JD2417970 - 33100). Since JD2433100 - see Table. Delta(V) var?

(Not logged in) » Add remark

**References**

Click reference title/citation to view in new window. Roll over index number to view submission details.

1 T.H. Astbury, 1905, JBAA 15, 270 1905JBAA...15..270A

Figure 17: Part of the AAVSO International Variable Star Index listing for RT Aur, showing Astbury listed as discoverer and, at the bottom, the reference to his JBAA discovery paper